\documentclass[twocolumn]{aastex63}
\usepackage[utf8]{inputenc}
\usepackage{comment}
\usepackage{amsmath}

\begin{document}
\title[Implications of a Temperature Dependent IMF III: Mass Growth and Quiescence]{Implications of a Temperature Dependent IMF III: Mass Growth and Quiescence} 

\author[0000-0003-3780-6801]{Charles L. Steinhardt}
\affiliation{Cosmic Dawn Center (DAWN)}
\affiliation{Niels Bohr Institute, University of Copenhagen, Lyngbyvej 2, K\o benhavn \O~2100, Denmark}

\author[0000-0002-5460-6126]{Albert Sneppen}
\affiliation{Niels Bohr Institute, University of Copenhagen, Lyngbyvej 2, K\o benhavn \O~2100, Denmark}
\affiliation{Cosmic Dawn Center (DAWN)}

\author{Hagan Hensley}
\affiliation{California Institute of Technology, 1200 E. California Blvd., Pasadena, CA 91125, USA}
\affiliation{Cosmic Dawn Center (DAWN)}

\author{Adam S. Jermyn}
\affiliation{Center for Computational Astrophysics, Flatiron Institute, New York, NY 10010, USA}

\author{Basel Mostafa}
\affiliation{California Institute of Technology, 1200 E. California Blvd., Pasadena, CA 91125, USA}
\affiliation{Cosmic Dawn Center (DAWN)}

\author[0000-0003-1614-196X]{John R. Weaver}
\affiliation{Niels Bohr Institute, University of Copenhagen, Lyngbyvej 2, K\o benhavn \O~2100, Denmark}
\affiliation{Cosmic Dawn Center (DAWN)}

\author{Gabriel Brammer}
\affiliation{Niels Bohr Institute, University of Copenhagen, Lyngbyvej 2, K\o benhavn \O~2100, Denmark}
\affiliation{Cosmic Dawn Center (DAWN)}

\author[0000-0003-3873-968X]{Thomas H. Clark}
\affiliation{California Institute of Technology, 1200 E. California Blvd., Pasadena, CA 91125, USA}
\affiliation{Cosmic Dawn Center (DAWN)}

\author{Iary Davidzon}
\affiliation{Niels Bohr Institute, University of Copenhagen, Lyngbyvej 2, K\o benhavn \O~2100, Denmark}
\affiliation{Cosmic Dawn Center (DAWN)}

\author[0000-0002-6459-8772]{Andrei C. Diaconu}
\affiliation{California Institute of Technology, 1200 E. California Blvd., Pasadena, CA 91125, USA}
\affiliation{Cosmic Dawn Center (DAWN)}

\author{Bahram Mobasher}
\affiliation{University of California, Riverside, 900 University Ave. Riverside, CA 92521}

\author[0000-0001-7633-3985]{Vadim Rusakov}
\affiliation{Niels Bohr Institute, University of Copenhagen, Lyngbyvej 2, K\o benhavn \O~2100, Denmark}
\affiliation{Cosmic Dawn Center (DAWN)}

\author{Sune Toft}
\affiliation{Niels Bohr Institute, University of Copenhagen, Lyngbyvej 2, K\o benhavn \O~2100, Denmark}
\affiliation{Cosmic Dawn Center (DAWN)}

\begin{abstract}
The stellar initial mass function (IMF) is predicted to depend upon the temperature of gas in star-forming molecular clouds.  The introduction of an additional parameter, $T_{IMF}$, into photometric template fitting, suggest most galaxies obey an IMF top-heavier than the Galactic IMF.  The implications of the revised fit on mass function, quiescence and turnoff are discussed. At all redshifts the highest mass galaxies become quiescent first with the turnoff mass decreasing towards the present. The synchronous turnoff mass across galaxies suggests quiescence is driven by universal mechanisms rather than by stochastic or environmental processes. 
\end{abstract}


\section{Introduction}
\label{sec:intro}

Several of the strongest constraints on galaxy formation and evolution are provided by measurements of the stellar mass function, or stellar mass distribution, of distant galaxies.  Perhaps the most longstanding is the discovery that the most massive galaxies complete their growth earliest, one of several related effects collectively termed downsizing \citep{Cowie1996,Fontana2006,Stringer2009,Fontanot2009}.  The existence of massive galaxies at high redshift requires high stellar baryon fractions \citep{Finkelstein2015}, or may even challenge the standard $\Lambda$CDM cosmological paradigm \citep{Steinhardt2016,Behroozi2018}.  The existence of massive quiescent galaxies at high redshift \citep{Toft2014,Glazebrook2017,Schreiber2018,Tanaka2019,Valentino2020} may pose a similar challenge. 

All of these results rely on measurements of galactic stellar masses.  Most of the light from a galaxy is emitted by only the most massive stars, comprising a small fraction of the full stellar mass.  The remainder of the stellar population must be inferred from assumptions which are difficult to test outside of our own Galaxy.  As a result, the luminosity of a distant galaxy can be determined far more precisely than its stellar mass. 

This is particularly true because of the advent of large photometric surveys which only have accompanying spectroscopy for a small fraction of their objects.    The Sloan Digital Sky Survey \citep{SDSSDR16} contains spectra for $\sim 0.5$\% of the $10^9$ objects with optical photometry.  The largest multi-wavelength survey with the infrared observations necessary to constrain stellar masses (cf. \citealt{Bradac2014}), COSMOS \citep{Scoville2007}, includes spectroscopy \citep{Hasinger2018} for approximately 2\% of the $10^6$ objects imaged \citep{Laigle2016}.  Ultra-deep surveys now contain $\gtrsim 10^3$ objects at $z > 6$ \citep{Bouwens2015,Bouwens2016,BUFFALO}, with very few spectra at these redshifts.  Thus, for the vast majority of galaxies, stellar mass determination comes from photometric template fitting, a set of techniques for fitting model spectra to photometry in order to determine redshift and physical parameters.

Photometric template fitting relies on the assumption that these models successfully describe galaxies.  Mathematically, there will always be a best-fit reconstruction for each galaxy over the allowed template space, regardless of whether the model space includes the true galaxy properties\footnote{Because it is known that these models do not entirely describe galaxies, goodness-of-fit metrics are typically not used to reject mismatches except in extreme cases.}.  However, if the model space is insufficient, that best-fit model will produce incorrect properties.  

The most critical assumption for stellar mass determination is the stellar initial mass function (IMF), or the distribution of stellar masses in a zero-age stellar population.  Because most of the stellar mass is inferred from only the rare, most massive stars, a small change in the shape of the IMF can produce a significant change in stellar mass.  A change in IMF will also change star formation rate, metallicity, age, extinction, and other inferred parameters.

The IMF has been historically assumed to be constant for all galaxies, as it can only be empirically measured in the Milky Way. That is, star formation in all galaxies, at all redshifts, is assumed to produce the same mass distribution that it does in the Milky Way. However, observational evidence also suggests that the IMF may not be universal \citep{Conroy2012,LaBarbera2013,Spiniello2014,Lyubenova2016,Lagattuta2017,vanDokkum2017}.  Further, physical models of star forming clouds strongly suggest that the distribution of stellar masses formed should depend on the temperature of the cloud, with a top-heavier IMF for higher temperatures \citep{LyndenBell1976,Larson1985,Bernardi2017,Jermyn2018}.  

Although direct observational measurements of gas temperatures are difficult, there is considerable indirect evidence to suggest that gas temperature vary.  Dust temperatures in star-forming galaxies are typically above 20 K, with higher dust temperatures found both towards high redshift and towards higher star formation rates (SFR) at fixed stellar mass and redshift \citep{Magnelli2014,Schreiber2018,Kokorev2021}.  Although luminosity-averaged dust temperatures are not guaranteed to be reliable indicators of gas temperatures in star-forming regions, the gas temperatures inferred from the methods used in this work lie in a similar range and have similar redshift dependence and SFR dependence (cf. \citet{Steinhardt2022a} for an extended discussion), although they are measured in entirely different ways.

\citet{Jermyn2018} developed a theoretical prescription for a one-parameter family of IMFs at some gas temperature $T_{IMF}$.  This work is the third in a series which studies the effects of fitting the COSMOS2015 \citep{Laigle2016} catalog with templates derived from those IMFs.  Paper I \citep{Sneppen2022} describes the fitting procedure, finding that nearly every galaxy is best fit with a $T_{IMF}$ higher than the Milky Way.  Thus, nearly every galaxy is best described with an IMF which is bottom-lighter (or top-heavier) than our own.  Paper II \citep{Steinhardt2022a} explores the implications of these results for star-formation and the star-forming ``main sequence''.  This work describes the effects on stellar mass functions and high-redshift cosmology.

The fitting procedure and dataset are summarized in \S~\ref{sec:methods}, with a full description given in \citet{Sneppen2022}.  The updated stellar masses derived from the updated catalog and resulting shift in mass functions are described in \S~\ref{sec:massfunc}.  These results imply a much stricter mass hierarchy for quiescence than previous results.  Whether the highest-redshift galaxies continue to produce tension with $\Lambda$CDM is described in \S~\ref{sec:highz}.  The broader implications of these results are discussed in \S~\ref{sec:discussion}.

\section{Methodology}
\label{sec:methods}

The template fitting presented here attempts to reproduce standard, existing techniques as closely as possible, with the sole exception of a single additional parameter which allows selection from a family of IMFs.  Full details on the technique, selection, uncertainties, covariances between parameters, and a comparison with fits using a fixed IMF are given in Paper I \citep{Sneppen2022}

Fits are performed on the largest multi-wavelength photometric dataset available, the COSMOS2015 catalog \citep{Laigle2016}. Objects are measured in as many as 26 filters: 12 broad bands (NUV, u, B, V, r, i, z, Y, J, H, and IRAC channels 1 and 2), two narrow bands (NB711 and NB816), and 12 intermediate bands.

The resulting catalog is fit using the photometric template fitting code EAZY \citep{Brammer2008}.  The standard version of EAZY fits galaxy SEDs as a linear combination of 12 basis templates, drawn themselves as linear combinations of 560 Flexible Stellar Population Synthesis (FSPS; \citealt{fsps}) synthetic spectra using different combinations of metallicity, extinction, and age but an identical IMF.  Here, equivalent basis sets are produced using the \citet{Jermyn2018} IMF,
\begin{equation} \xi(m) \propto \begin{cases} m^{-0.3} & m < 0.08M_{\odot}\left(\frac{T_{IMF}}{T_0}\right)^2\\
          m^{-1.3} & 0.08M_{\odot}\left(\frac{T_{IMF}}{T_0}\right)^2 < m < 0.5M_{\odot}\left(\frac{T_{IMF}}{T_0}\right)^2\\
      m^{-2.3} & m > 0.5M_\odot \cdot \left(\frac{T_{IMF}}{T_0}\right)^2 \end{cases} \end{equation}

Where $T_0$ = 20 K, so at $T_{IMF} = T_0$, this produces a standard Kroupa IMF \citep{Kroupa2001}. For each IMF, a set of 560 FSPS templates is constructed corresponding to the same combinations of age, star formation history, extinction, and metallicity as in the standard EAZY library.  Those are then reduced to 12 basis templates, again using the same procedure as for the standard EAZY library.  Fits are performed over a grid of IMFs, spaced every 1K for $8\textrm{K} \leq T_{IMF} \leq 60\textrm{K}$.  

Thus, $T_{IMF}$ determines the top-heaviness of the IMF, with the variability of the IMF causally interpreted as being set by temperature in star forming clouds. Given other theoretical prescriptions other dependencies on temperature have been suggested, but the presented approach is agnostic to the theoretical model. A translation to other mass scaling relations is seen in \cite{Sneppen2022}. Additionally, it should be noted that the IMF is derived from fitting the existing stellar population. Thus, $T_{IMF}$ does not probe the gas temperature in star-forming clouds at the time the observed light was emitted, but instead when the existing stellar population was formed.    

The value of $T_{IMF}$ yielding the minimum reduced $\chi^2$ is chosen as the best-fit temperature for each object.  For most of the high signal-to-noise objects, there is a single local $\chi^2$ minimum within the grid, which is also the global minimum.  For the remainder of the catalog, mostly objects with fewer bands or low signal-to-noise, there are multiple local minima or the only local minimum lies at 8K or 60K.  These objects are discarded from the final catalog, with the exception of the population described in \ref{subsec:twobreaks}.

For each object at any $T_{IMF}$, EAZY outputs a best-fit and a photometric redshift $z_{phot}$ and a linear combination of basis templates to the observed photometry.  Typically the photometric redshift is similar for all $T_{IMF}$, as it is mainly constrained by spectral breaks rather than by the shape of the SED.  Galactic parameters including stellar mass, SFR, metallicity, and age are then constructed as a (luminosity-weighted) linear combination of the basis templates.  For the comparisons shown in this work, star-forming and quiescent galaxies are separated using a UVJ diagram. Rest-frame U, V, and J fluxes are calculated from the best-fit reconstructed spectrum, shifted to the rest-frame using the calculated photometric redshift and integrated over the relevant filters.

As described in detail in Paper II, most galaxies in COSMOS are best fit with $T_{IMF} \gtrsim 25 K$, as compared with a Galactic IMF at $T_{IMF} = 20$K.  Further, the best-fit stellar mass, SFR, and other parameters are sensitive to changes in $T_{IMF}$ on this scale (\cite{Sneppen2022}, Fig. 8).  Thus, the best-fit properties of most objects in COSMOS change significantly following the introduction of $T_{IMF}$ into the fitting process.  The remainder of this paper focuses on the ways in which this alters stellar mass functions, and the implications for our understanding of stellar mass growth and quiescence.

\subsection{Well-Measured Objects with Multiple Minima}
\label{subsec:twobreaks}

\begin{figure}[ht]
    \includegraphics[width=0.45\textwidth]{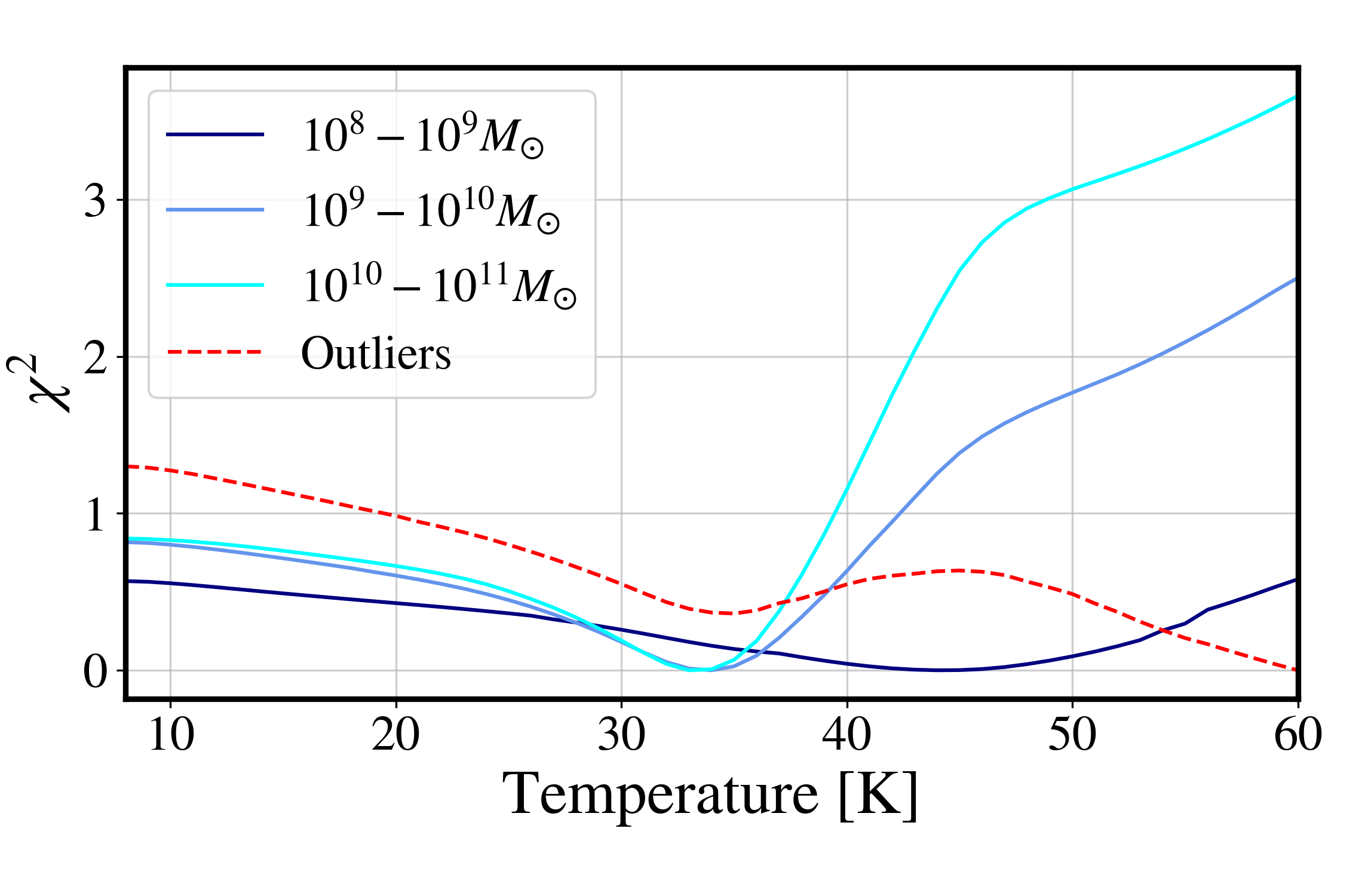}
    \caption{Average $\Delta \chi^2(T_{IMF})$ at various masses for star-forming galaxies at $0.95 < z < 1.05$.  At every mass, there is a clear local minimum at a similar $T_{IMF}$.  In addition, the slope changes with mass, and a small outlier population of the most massive star-forming galaxies galaxies (red, dashed) are best fit with an unphysically high $T_{IMF}$.  A plausible explanation is that the local minimum between 30K and 35K indicates gas temperatures in star-forming regions.  The extremely high $T_{IMF}$ would then indicate an additional breakpoint in the stellar mass distribution coming from a lack of very high mass stars, a possible indication of turnoff.}
    \label{fig:stackminima}
\end{figure}

Most objects without a single, clear global minimum in $\chi^2(T_{IMF})$ have a complex landscape due to large measurement uncertainties and correspondingly poorly-constrained fits.  However, there is also a small class of well-measured objects, typically high-mass star-forming galaxies, without a single global minimum.  These objects typically exhibit a general trend of an improved fit towards high, physically unreasonable values of $T_{IMF}$.  In addition, there is a local $\chi^2$ minimum at a value of $T_{IMF}$ similar to other star-forming galaxies at the same redshift.  A stacked $\chi^2$ landscape (Fig. \ref{fig:stackminima}) shows the change in overall slope of $\chi^2(T_{IMF})$ towards high mass.  
This implies a combination of two effects, one common to all star-forming galaxies and a second preferentially occurring at high stellar mass.

A change in $T_{IMF}$ results in an IMF of similar shape to a Kroupa IMF, but with break masses in different locations.  Because starlight is dominated by high-mass stars, in effect, measurements of $T_{IMF}$ are predominantly measurements of the higher-mass break mass.  Finding that star-forming galaxies have a similar $T_{IMF}$ essentially indicates that they have similar breaks in their stellar mass distribution.

However, changes in the stellar population can be produced not only by a change in the IMF, but also by a change in the star formation history (SFH).  In principle, the IMF and SFH are entirely degenerate\footnote{A particularly poor choice of IMF could be ruled out because producing the observed stellar population would require a the SFH to be negative at some times.  However, given the other significant uncertainties, a statistically significant rejection of a plausible IMF is unlikely.}.  Given any choice of IMF, a SFH can be constructed to match any stellar population.  The highest-mass stars with the shortest lifetimes can be used to determine the recent SFR.  The IMF is then used to construct a full stellar population at that SFR.  Subtracting this population, the next highest-mass objects in the remaining population are then used to construct the SFR at an earlier time.  Iterating will produce a binned SFH which can be combined with the assumed IMF to construct the present stellar population.  This approach has been used to determine extragalactic SFHs \citep{Panter2003,Sanchez2019}, assuming a Galactic IMF.

Here, the stellar population is decomposed by allowing the IMF to vary rather than the SFH.  The standard EAZY methodology assumes a delayed-tau SFH \citep{Brammer2008}, with the stellar population ages in this work allowed to range from 0.02-20 Gyr \citep{Sneppen2022}.  Thus, a high-mass break in the stellar population is assumed to correspond to a high-mass break in the IMF, and therefore a very high $T_{IMF}$.  However, the same lack of very high-mass main sequence stars can also be produced by a rapid decline in SFR in the recent past, with the time since that decline corresponding to the main sequence lifetime at the break mass.  In that case, instead of a high $T_{IMF}$, such a break would be a possible indicator of very recent star formation turnoff.

Such an interpretation would fit nicely with the additional evidence presented in \S~\ref{subsec:turnoff}.  For the remainder of this work, rather than excluding these objects from the catalog, the choice is made to include them using the best-fit properties at the local minimum of $T_{IMF}$ between 20K and 45K.  A choice to use the far higher value of $T_{IMF}$ would result in significantly lower stellar mass estimates.  Thus, either using the higher value of $T_{IMF}$ or excluding this sample entirely would only strengthen the conclusions regarding high-mass turnoff in the following sections.  However, because this is only a small population, those conclusions are not sensitive to this choice.

\section{Effects on Stellar Masses and Quiescence Mechanisms}
\label{sec:massfunc}

\subsection{Effects of IMF ``Temperature'' on Stellar Masses}

\begin{figure}[ht]
    \centering
    \includegraphics[width=0.45\textwidth]{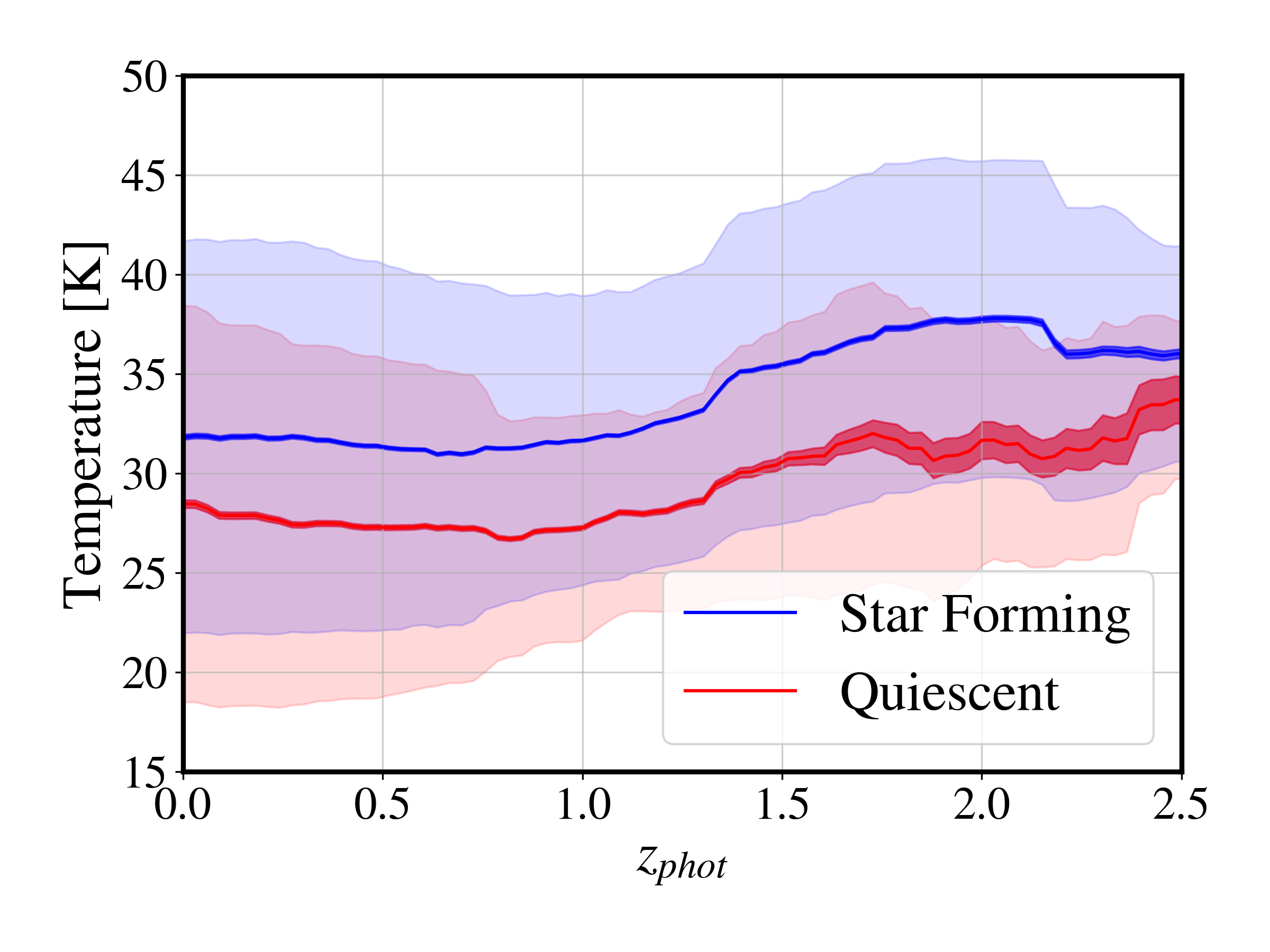}
    \caption{Mean best-fit $T_{IMF}$ for quiescent (red) and star-forming (blue) galaxies as a function of redshift. Light shading indicates the 1$\sigma$ spread of individual observations, with dark shading indicates the uncertainty on the mean.
    Quiescent galaxies lie at temperatures cooler than their star-forming counterparts, but both populations typically require top-heavier IMFs than a Galactic IMF. The Anderson-Darling test-statistic quantifies that within any redshift-bin of width $\Delta z=0.25$ from $z=0$ to $z=2$) the two distributions yield a p-value of less than $10^{-10}$ to be drawn from the same underlying distribution, and a p-value $<0.1$\% from the smaller sample at $z \sim 2.5$. At higher redshifts the mean temperature of both populations increases.}  
    \label{fig:imftemps}
\end{figure}

The best-fit $T_{IMF}$ at nearly every redshift is not $20$K, corresponding to a Galactic IMF, but rather higher (Fig. \ref{fig:imftemps}).  
It might be expected that a top-heavier IMF will always produce a lower stellar mass, since massive stars have a smaller mass-to-light ratio, and therefore a distribution with more massive stars also has a smaller mass-to-light ratio. However, changing the temperature of the fit might also change other parameters that are degenerate with the stellar mass (as detailed in Paper I). Thus, it is first necessary to recompute individual stellar masses from templates for the objects with sufficient signal to noise to constrain them.  Subsequently,  the effects on the entire stellar mass distribution can be estimated from those best-fit stellar masses in order to correctly understand how that distribution changes with the introduction of $T_{IMF}$.

\subsection{Uncertainties and Eddington Bias}

The resulting stellar mass distribution exhibits many of the properties reported in previous studies which assumed a Galactic IMF.  However, stellar masses determined from template fitting have large uncertainties.  It is therefore necessary to model this uncertainty in order to determine the true, underlying stellar mass function (SMF) from the observed stellar mass distribution.  

The observed SMF can be thought of as the true SMF convolved with some error function, changing the shape of the SMF. Since what is measured is the distribution of objects above the stellar mass completeness cut, this creates a high-mass bias in observed SMFs, which is a form of the Eddington bias.  Here, this bias is corrected using the methodology presented in \citet{Ilbert2013}. Assuming that the true SMF follows a Schechter function and the scatter follows the same distribution determined by \citet{Ilbert2013}, a best-fit error-convolved Schechter function is calculated for the observed SMF, which can then be easily deconvolved to obtain an estimate of the true SMF. In practice, the effect of this error convolution is primarily that the observed mass function is overestimated at the high-mass tail, which deconvolution corrects.  Thus, the correction will appear most significant at high redshift, as the detection threshold becomes closer to the maximum observed mass. 

\subsection{Revised Mass Completeness Cut}
Constraining an additional parameter ($T_{IMF}$) as compared with previous fits requires additional information, and therefore stricter SNR-cut (detailed in Paper I).  This cut will affect the stellar mass distributions observed, as it rejects less luminous and therefore typically less massive galaxies. Thus, the mass distribution is no longer complete to the same degree at the same masses as in previous COSMOS studies \citep{Davidzon2017}. Therefore, for further analysis the mass completeness cut is reevaluated at all redshifts. 

The mass completeness is estimated at fixed redshift by determining the highest masses observed at the flux in K-band which separates cut and accepted objects. A threshold of $K=0.5 \mu $Jy yields a clear delimiter between the cut and accepted object for all redshifts with a $>$95 \% true positive rate and $<$10 \% false negative rate. 

Because $T_{IMF} > 20$K, the inferred stellar masses have decreased.  Thus, at every redshift, the typical stellar mass and thus the stellar mass distribution in this work lies below that of previous studies.  Thus, there are two effects which change the mass completeness in opposite directions.  The requirement for higher signal-to-noise leads to a brighter completeness limit.  However, the higher $T_{IMF}$ means that the same flux comes from a lower inferred stellar mass.  In practice, the former effect is more significant, and therefore using a variable IMF, the stellar mass completeness threshold is higher at most redshifts.

\subsection{Turnoff Fractions and Ordering}

The best-fit {$T_{IMF}$} for star-forming galaxies is higher than for quiescent galaxies at every redshift.  As a result, the decrease in stellar mass compared with previous measurements is also greater for star-forming galaxies. This shift in the relative masses of star-forming and quiescent galaxies might help to explain a puzzling feature of previous turnoff studies at high redshift.  The typical growth of star-forming galaxies is monotonic in mass; increasing mass at fixed redshift results in higher SFR but lower sSFR \citep{Noeske2007,Peng2010,Speagle2014,Steinhardt2014a,Schreiber2018}.  The same is true of hierarchical merging, in which smaller halos virialize before larger ones (cf. \citealt{Press1974,Sheth2001}) and of several results which have been described as downsizing \citep{Cowie1996,Fontana2006,Stringer2009,Fontanot2009}, showing that in several ways more massive galaxies evolve more rapidly than their less massive counterparts.

However, the quiescent fraction of galaxies as found using a Galactic IMF instead shows that the first galaxies to turn off, at $z > 2$, lie in the middle of the mass distribution, around $10^{10.5} M_\odot$ (\citet{Davidzon2017}, Fig. 12).  At the same redshift, nearly all galaxies at both higher and lower stellar masses are still star-forming.  
Further, the star-forming main sequence appears to be universal \citep{Steinhardt2014b}, with nearly all star-forming galaxies at the same stellar mass growing at the same rate at any given time.  If turnoff is a natural result of the process which drive star formation, by the same principle one should expect turnoff to occur at a similar time in an ensemble of galaxies of a given stellar (or halo) mass at any fixed redshift.  Thus, at most masses, nearly none or nearly all galaxies should be quiescent, with the mass range currently undergoing turnoff being the sole exception.  However, studies using a Galactic IMF instead find that a mixture of star-forming and quiescent galaxies at the same stellar mass is most typical \citep{Ilbert2013, Muzzin2013, Davidzon2017}.  For example, the quiescent fraction of $M \sim 10^{11} M_\odot$ galaxies lies between 20\% to 80\% from $z=4$ to $z=0.8$ in those studies.

At lower redshift, however, there is a different, far simpler picture.   The highest-mass galaxies turn off first, and at every redshift, nearly all of the most massive galaxies have become quiescent.  Lower-mass galaxies are predominantly star-forming.  This result follows expectations from observations of the star-forming main sequence and mass downsizing.  It suggests that turnoff is a natural extension of a galaxy's evolution as it exits the high-mass end of the star-forming main sequence, rather than an independent event which can happen to galaxies at any stage of their lifetimes. 

Using the best fit from a family of IMFs strengthens this picture.  Quiescent galaxies have an a nearly-Galactic IMF, so that their best-fit stellar masses will be similar to previous studies.   However, star-forming galaxies have higher $T_{IMF}$, with $T_{IMF}$ further increasing towards higher redshift \citep{Steinhardt2020}.  Thus, the best-fit stellar masses for star-forming galaxies will be systematically lower than in previous studies.  Further, they will systematically shift relative to the quiescent population.  Thus, the quiescent fraction will increase at high mass.

\begin{figure*}[ht]
    \centering
    \includegraphics[width=\linewidth]{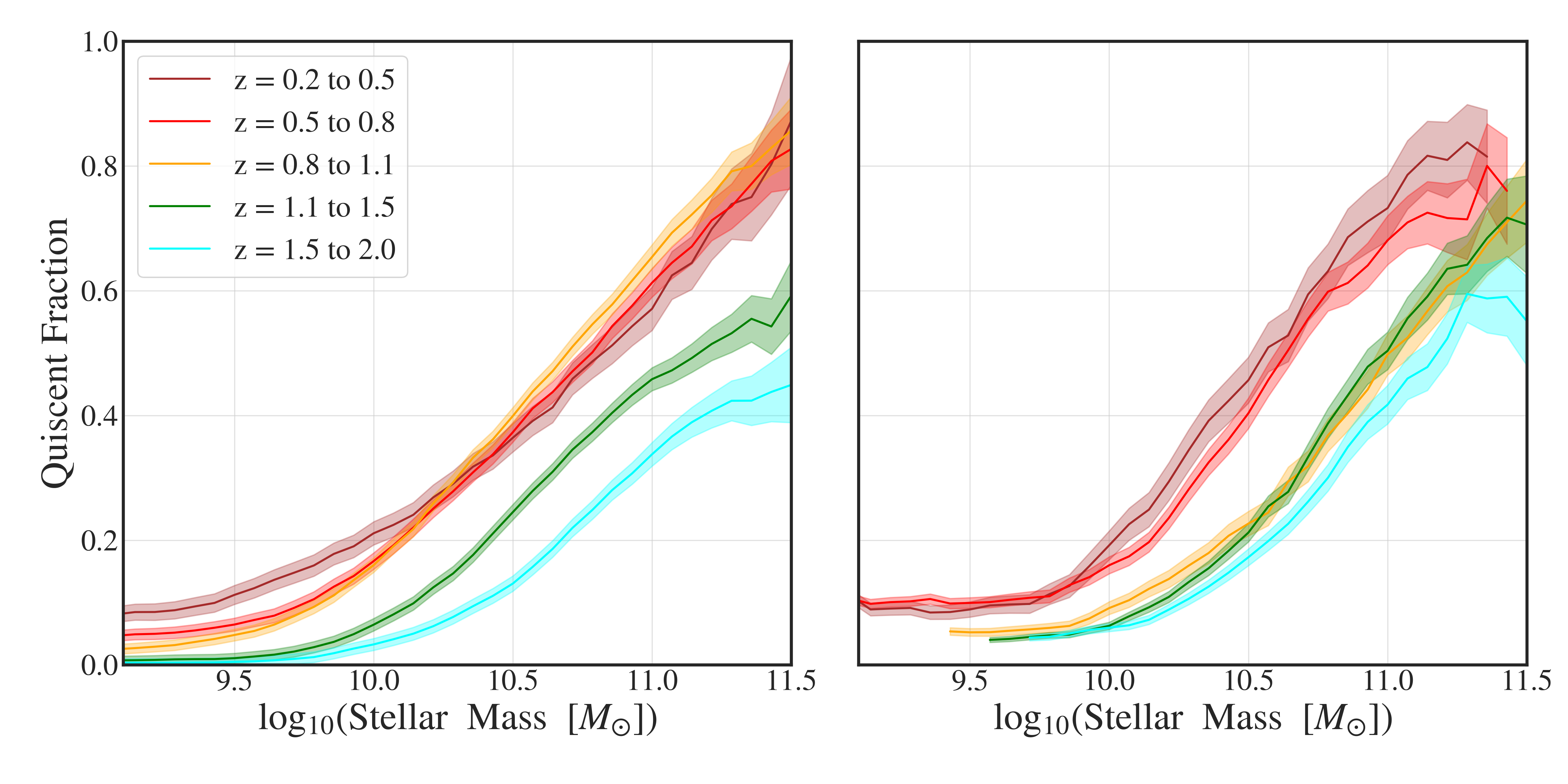}
    \caption{A comparison between the quiescent fractions at $z < 2$ using the \citet{Laigle2016} COSMOS2015 catalog (left), which uses a Galactic IMF for all fits, and the catalog in this work (right).  For both catalogs, the most massive galaxies at each redshift are nearly all quiescent, and the least massive are nearly all star-forming.  Both panels only show objects for which $T_{IMF}$ can be constrained, which restricts this to the $\sim 10$\% brightest objects in K band.  Too few objects at $z > 2$ are bright enough to constrain $T_{IMF}$, so the quiescent fractions at those redshifts cannot be measured directly using the techniques in this work.}
    \label{fig:qfraccomp}
\end{figure*}

Within the range covered by this study ($z \lesssim 2$), this produces an even sharper picture than using the \citet{Laigle2016} catalog on the same dataset (Fig. \ref{fig:qfraccomp}).  At all redshifts where the quiescent fraction can be measured directly allowing a best-fit IMF, the most (stellar) massive galaxies at each redshift are nearly all quiescent and the least massive nearly all star-forming.

\subsubsection{Extrapolation to Higher Redshifts}

\begin{figure}[ht]
    \centering
    \includegraphics[width=\linewidth]{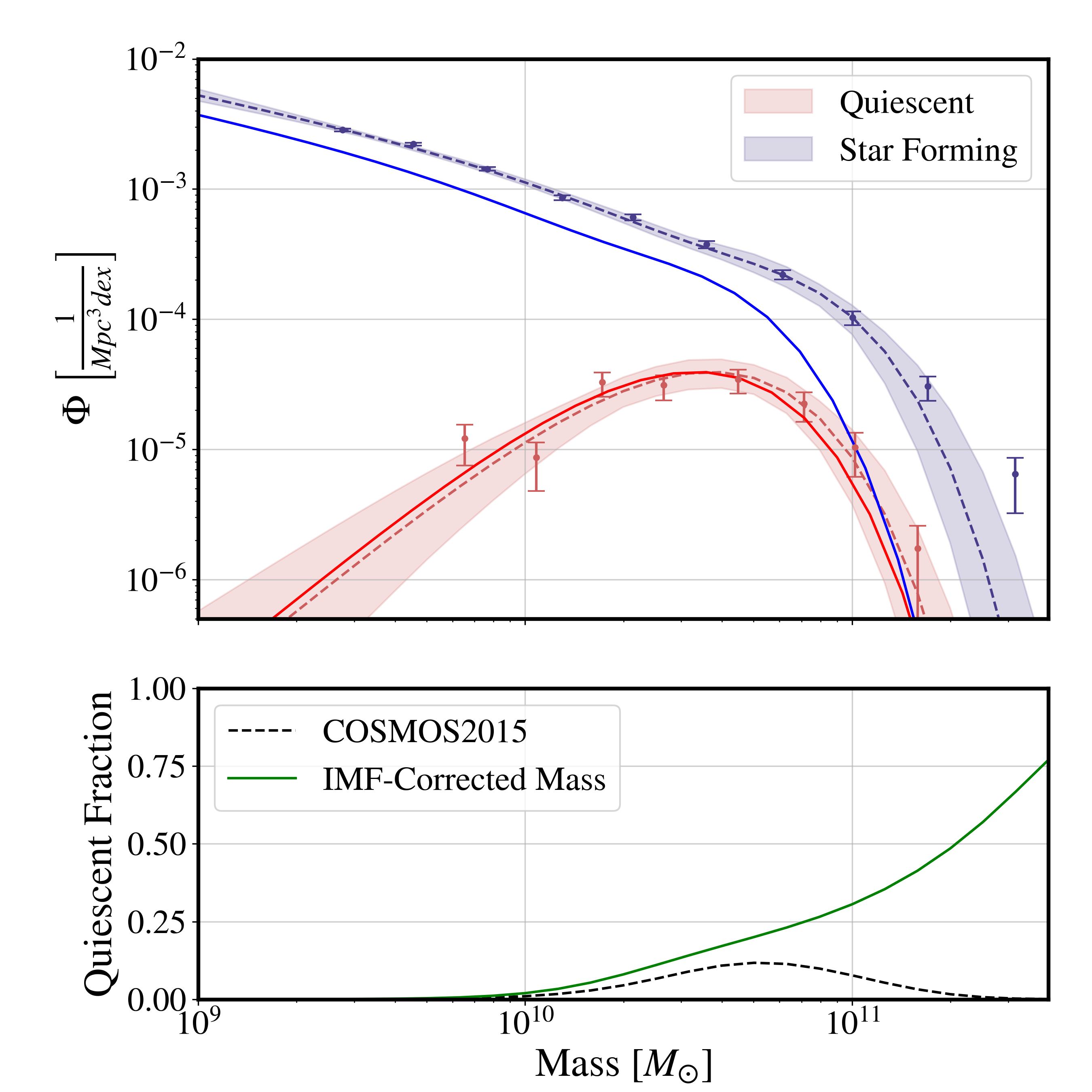}
    \caption{Top: Galaxy Stellar Mass Functions in range $2.5<z<3$ for quiescent and star-forming population as reported in \cite{Davidzon2017} (dashed lines) and given a shifted IMF (fully drawn lines). Bottom: The corresponding COSMOS2015 (dashed black) and IMF-varied (green line) quiescent fraction. Notable correcting for $T_IMF$ suggest a monotonic increase with stellar mass of the quiescent fraction.}
    \label{fig:extrapolation}
\end{figure}

At $z > 2$, COSMOS2015 presents an entirely different picture of quiescence than both COSMOS2015 and the variable-IMF fits at $z < 2$.  The \citet{Davidzon2017} fits find that the most massive galaxies at $2.5 < z < 3.0$ are star-forming (reproduced in Fig. \ref{fig:extrapolation}), and that the quiescent fraction peaks at $\sim 10^{10.5}M_\odot$, in the middle of the distribution.  As described above, this result is difficult to reconcile with the star-forming main sequence, with downsizing.  So, it is natural to consider whether using a best-fit IMF might solve this ``problem'' by reducing star-forming stellar masses more than quiescent ones and yielding the more intuitive result found at lower redshifts.

COSMOS2015 has a large enough sample to constrain the quiescent fraction out to $z = 3$ without additional assumptions\footnote{Although quiescent fractions at $z > 3$ are also reported, they rely on tying fit parameters to lower-redshift galaxies, which is likely a bad assumption with a redshift-dependent $T_{IMF}$.} However, there are few quiescent galaxies at those redshifts with sufficient signal to noise to constrain the IMF, and thus the quiescent fraction cannot be measured at those redshifts using the techniques in this paper.  Instead, it is necessary to extrapolate to these redshifts based on a comparison between stellar masses using Galactic and best-fit IMFs at lower redshifts.  

Ideally, one would start by comparing the COSMOS2015 masses with those in this work.  A typical offset could be measured at every $T_{IMF}$, perhaps separately for quiescent and star-forming galaxies if the offset is different at the same temperature for young and old stellar populations.  Extrapolating the characteristic $T_{IMF}$ (cf. \citealt{Sneppen2022}, Fig. 7) for star-forming and for quiescent galaxies out to $z = 3$, those temperatures and corresponding IMFs could then be used to produce offsets and adjust the \citet{Davidzon2017} stellar mass functions for each population.  

However, in addition to a change in the best-fit IMF, there are three other differences which can significantly alter inferred stellar masses.  First, COSMOS2015 catalog is based on the LePhare photometric template fitting code, and this work uses EAZY.  A comparison of LePhare and EAZY stellar masses for the COSMOS2020 catalog shows a $\sim 0.2$ dex offset \citep{Weaver2022}.  Second, COSMOS2015 fits galaxies with a Chabrier IMF, and the 20K IMF used here is instead identical to a Kroupa IMF.  This, too, predominantly produces a systematic shift in the inferred stellar mass, although in principle there could also be a slight dependence on the age of the stellar population.  Finally, a small fraction of objects are best fit with significantly different redshifts, due to reinterpretation of a Lyman break as a Balmer break, or vice versa.  None is these is the effect of interest here.  However, they potentially dominate a direct comparison of masses between COSMOS2015 and this work. 

Fortunately, these other differences should affect both star-forming and quiescent galaxies similarly, whereas a change in $T_{IMF}$ does not.  Thus, in order to determine the relative shift between star-forming and quiescent mas functions, it is sufficient to determine the relative shift due to the change in $T_{IMF}$ alone.  Here, that shift is modeled as follows at each redshift:
\begin{enumerate}
    \item {The characteristic $T_{IMF}$ is determined separately via linear extrapolation for star-forming and quiescent galaxies at $2.5 < z < 3.0$.  This produces $T_{IMF} = 35$K for star-forming galaxies and $25$K for quiescent ones.}
    \item {The median offset between the best-fit stellar mass at that $T_{IMF}$ and the stellar masses for those galaxies using a 20K IMF is found (ie. the offset illustrated in Fig. 9 in \cite{Sneppen2022}).}
    \item{Those offsets are then applied to the \citet{Davidzon2017} stellar mass functions at $2.5 < z < 3.$}
\end{enumerate}

The results of applying those offsets to the \citet{Davidzon2017} (Fig. \ref{fig:extrapolation}) shows a significant qualitative difference.  Whereas the \citep{Davidzon2017} quiescent fraction peaks in the middle of the distribution, this extrapolation produces a result similar to the low-redshift answer.  The highest-mass galaxies turn off first, and all of the most massive galaxies have become quiescent.  Lower-mass galaxies are predominantly star-forming.  
Thus, the effects of allowing a best-fit IMF are potentially sufficient to restore the more intuitive, low-redshift answer.  

If the low-redshift shape of the quiescent fraction is indeed the shape at all redshifts, a further implication is that turnoff appears to be a one-time, permanent process rather than galaxies having several shorter periods of quiescence between rounds of star formation.  Alternating periods of quiescence and growth would produce a lower-mass quiescent population, rather than having the simpler result that only the most massive galaxies are found to be quiescent.  A one-time, permanent turnoff as suggested by the variable-IMF fits is easier to reconcile with both the star-forming main sequence and downsizing, as it would simply means that galaxies which all grow in the same way naturally turn off at the end of that growth. However, these results still do not suggest any specific physical mechanism for turnoff.

\subsection{Turnoff Mass and Quiescent Galaxy Selection}

Following the introduction of $T_{IMF}$, virtually all of the highest-mass galaxies at fixed redshift are quiescent and virtually all of the lowest-mass galaxies are star-forming, with a sharp transition between the two.  This transition is so sharp that it even provides an approximate selection method for quiescent galaxies: simply use the best-fit stellar mass, assuming that the highest mass galaxies are most likely to be quiescent.

This is a far less sophisticated method than ones in current use.  Those include color selection criteria \citep{Williams2009,Arnouts2013}, template fitting followed by selection of galaxies with low sSFR \citep{Brammer2008,Laigle2016}, and most recently, a variety of machine learning approaches \citep{Leja2019,Davidzon2019,Steinhardt2020,Shahidi2020,Turner2021}.  With the exception of color selection, which is boolean, the other criteria instead provide a score or ranking, which can then be turned into a probability that the galaxy is quiescent.  For example, in template fitting, a lower sSFR corresponds to a higher likelihood of quiescence.  A threshold is then typically selected, with objects above the threshold considered quiescent and below the threshold star-forming.

\begin{figure}[ht]
    \centering
    \includegraphics[width=0.5\textwidth]{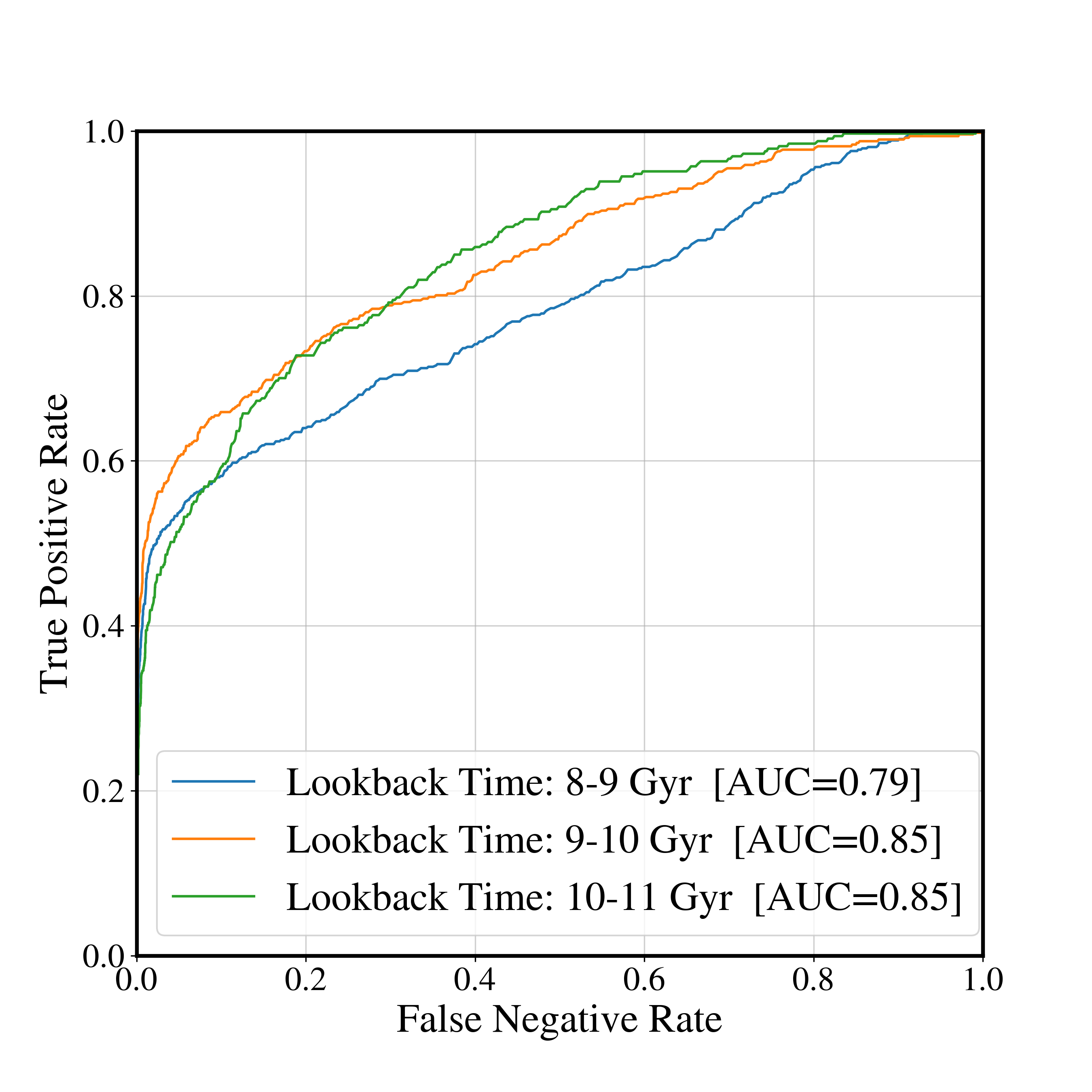}
    \caption{Receiver operating characteristic (ROC) curve using stellar mass as a proxy for likelihood of quiescence.  Although this is a far cruder selection than several methods in current use, the overall $\Sigma$ROC (or AUC) of $\sim 0.8$ at a wide range of redshifts is competitive with several current techniques, indicating the surprising sharpness and completeness of the transition from lower-mass star-forming galaxies to higher-mass quiescent ones.}
    \label{fig:roc}
\end{figure}

The success of these criteria is often evaluated using a receiver operating characteristic (ROC) curve (Fig. \ref{fig:roc}), which evaluates the tradeoff between false positive and false negative rates at different thresholds.  

As a single, summary metric, $\Sigma$ROC\footnote{This is typically called AUC, or area under the curve, in machine learning literature.} is a rank sum test, similar to the Mann-Whitney U Test.  Although stellar mass is a far cruder selection than several methods in current use, the overall $\Sigma$ROC (or AUC) of $\sim 0.8-0.85$ at a wide range of redshifts is competitive with several current techniques, indicating the surprising sharpness and completeness of the transition from lower-mass star-forming galaxies to higher-mass quiescent ones.

This sharpness also makes it natural to define a single transition, or turnoff mass, $m_{to}$, between the two populations.  Because the transition takes place in a relatively narrow mass range, most definitions of $m_{to}$ will provide similar results.  Here, $m_{to}$ is defined in every redshift as the mass at which 50\% of galaxies are quiescent in a smoothed fit.  

\begin{figure}[ht]
    \centering
    \includegraphics[width=0.5\textwidth]{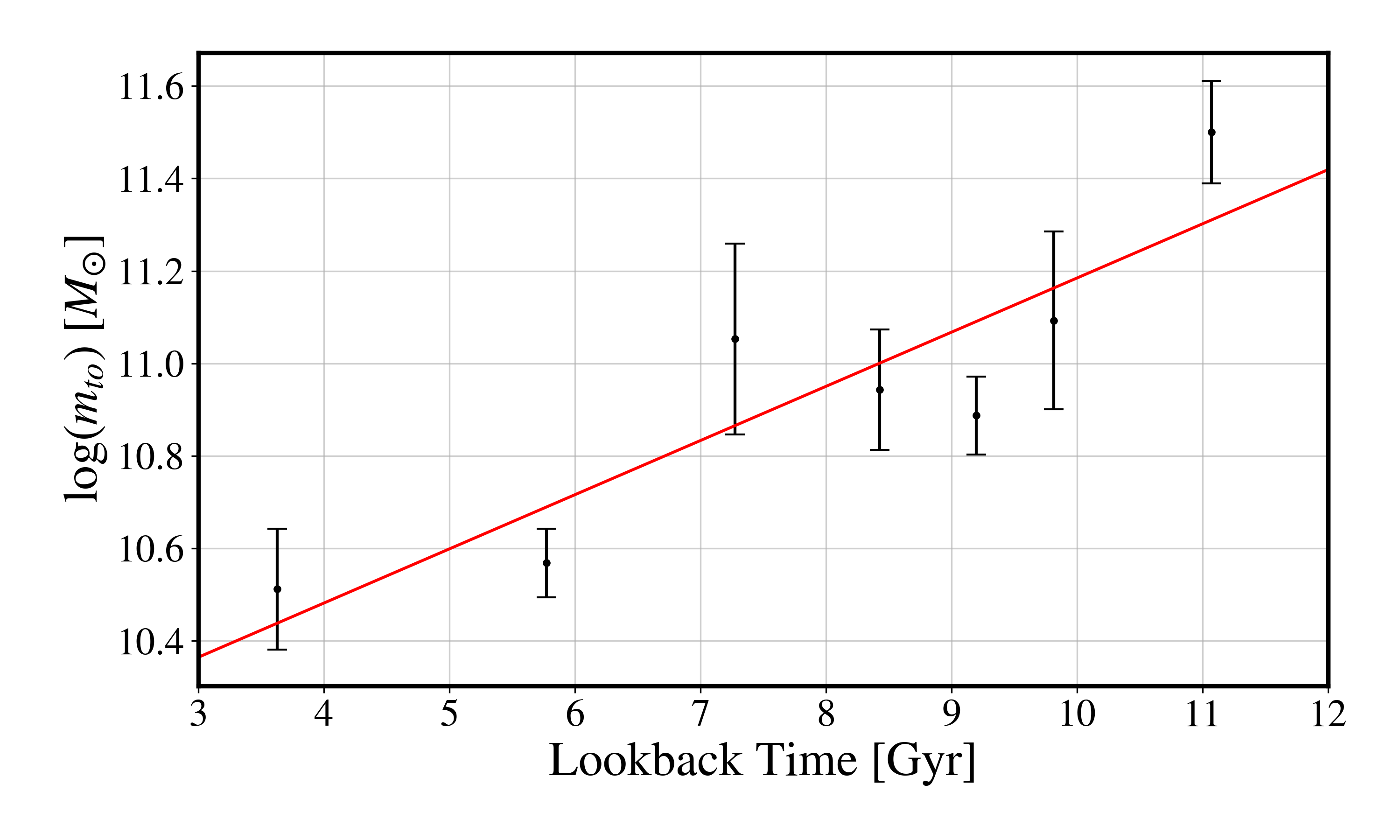}
    \caption{Turnoff mass $m_{to}$ as a function of lookback time.  $\log m_{to}$ is reasonably well approximated as linear in lookback time, corresponding to an exponential $m_{to} \propto e^\tau$. Red line indicates best fit: $\log (m_{to}/M_\odot) = (10.11 \pm 0.15) + (0.11 \pm 0.02) \tau $, $\chi^2/dof=10.5/5$ }
    \label{fig:turnoffmass}
\end{figure}

The resulting $m_{to}$ decreases towards lower redshift (see Fig. \ref{fig:turnoffmass}).  

Further, $\log m_{to}$ is nearly linear in lookback time, with best-fit $\log (m_{to}/M_\odot) = (8.57 \pm 0.18) + (0.26 \pm 0.03) \tau $.  This linearity of $\log m_{to}$ in lookback time is reminiscent of a similar linearity in $\log \textrm{sSFR}$ along the star-forming main sequence \citep{Speagle2014}.  This is further support for the possibility that quenching might not be an external event that ends star formation, but rather should be though of as the natural endpoint of the same feedback mechanisms which drive the star-forming main sequence.

\section{Discussion}
\label{sec:discussion}

Most of the inferred properties for galaxies in previous studies have relied on the assumption that a Galactic IMF is Universally applicable.  However, when photometry is fit to models incorporating a family of possible IMFs, nearly every galaxy outside the local Universe is instead best fit with one top-heavier than the Milky Way.  This top-heaviness is best described in terms of a single parameter, $T_{IMF}$, which is hoped to correspond to a typical gas temperature in star-forming clouds.  A higher $T_{IMF}$ corresponds to a top-heavier, or more accurately bottom-lighter, IMF.

Although in aggregate galaxies exhibit $T_{IMF}$ ranging from $\sim 20-45$K, at any fixed redshift the distribution is far narrower.  Star-forming galaxies are best fit with a characteristic $T_{IMF}(z)$ which rises from $\sim 25$K at $z \sim 0$ to nearly $35$K by $z = 2$.  Quiescent galaxies have a lower characteristic temperature, which exhibits either only weak or negligible redshift dependence.

\subsection{The Most Massive, Highest-Redshift Galaxies}
\label{sec:highz}

Because $T_{IMF}$ is generally largest for the highest-redshift galaxies, the extrapolated differences between the stellar masses shown here and previous results using Galactic IMFs are also most significant at high redshift.  This is of particular interest because the most massive galaxies as measured using Galactic IMFs are sufficiently massive to challenge $\Lambda$CDM \citep{Steinhardt2016,Behroozi2018}. 

The fits in this work produce lower stellar masses at high redshift.  However, because the most massive star-forming galaxies at any redshift exhibit lower best-fit $T_{IMF}$ (see Paper II), the decrease in stellar mass is smaller for the most massive galaxies.  As a result, there is still a discrepancy between the theoretical halo mass function and the stellar mass function which would result from a redshift-independent stellar mass-halo mass relation.

\subsection{Recalibrating Simulations}
\label{sec:abundancematching}

One of the conclusions of this work is that the observed stellar mass function at high redshift differs significantly from previous studies. 
Because prior mass functions were well-reproduced by calibrated simulations of galaxy formation within halos (cf. \citealt{Vogelsberger2020}), this suggests that a new calibration may be necessary.

Such a calibration has not yet been done, although given the large number of free parameters in simulations it seems most likely that such a calibration is possible.
Should it prove impossible to reproduce our stellar mass functions with calibrated simulations within $\Lambda$CDM, that would be good evidence of a new tension between cosmology, template-fit galactic properties, and numerical simulations.

\subsection{Quiescence Mechanisms}

This difference in quiescent and star-forming IMFs is, at a minimum, yet another indication of the well-studied bimodality in observed galaxy properties, separating star-forming galaxies from quiescent ones.  However, it should be stressed that unlike most temperature measurements, the best-fit IMF is backward-looking.

Like most properties determined from photometry, $T_{IMF}$ measures a luminosity-weighted average.  In practice, galaxies likely are comprised of a mixture of stellar populations formed at a range of different $T_{IMF}$, given that the typical star-forming $T_{IMF}$ is redshift-dependent.  So, describing galaxies with a single $T_{IMF}$ is analogous to describing them with a single best-fit age for their stellar population.\footnote{It is also quite possible that different star-forming clouds within the same galaxy at the same redshift may lie at a range of temperatures.  If so, the true IMF for that galaxy may not correspond to any specific $T_{IMF}$, but rather a weighted sum of different IMFs which the fitting routine will approximate as a single IMF corresponding to a single, best-fit $T_{IMF}$.}

In both cases, these luminosity-weighted averages are dominated by the same population of the most massive stars still on the main sequence.  Typical age measurements indicate that the luminosity-weighted stellar population is typically $\sim 10^8$yr old in a star-forming galaxy and far older in a quiescent one.  Thus, $T_{IMF}$ for a star-forming galaxy describes the IMF $\sim 10^8$yr earlier than when the light is emitted, and $T_{IMF}$ for a quiescent galaxy describes the IMF during its last significant epoch of star formation.

As a result, the bimodality between the behavior of star-forming and quiescent $T_{IMF}$ does not merely indicate that the two populations are distinct.  Rather, it shows that quiescent galaxies had already taken on distinct properties from star-forming ones, including a more Milky Way-like IMF, {\em while they were still star-forming}.   

One possibility is that this might allow a search for quenching, rather than quenched, galaxies.  Because the strong change in galaxy color from blue to red only happens $\gtrsim 500$Myr after turnoff \citep{Wild2020}, color selection only finds galaxies which have long since quenched.  If star-forming galaxies begin to exhibit distinct properties even while still forming stars, it might be possible to select these galaxies while they are still quenching, providing far more information about possible mechanisms.  A simple selection proposed here is to look for galaxies with star-forming colors but low $T_{IMF}$ more characteristic of quiescent galaxies.  Further work might develop improved selection criteria.

The drop in $T_{IMF}$ also indicates that galaxies are actively forming stars at a high rate even as they begin quenching.  If $T_{IMF}$ truly reflects gas temperature in star-forming regions, it also helps to distinguish between different possible mechanisms.  For example, AGN heating would quench star formation by heating gas in star-forming regions, inhibiting the collapse into stars \citep{DiMatteo2005}.  For the stars which are still able to form, they would be expected instead form under warming conditions, so that quenching galaxies should instead exhibit $T_{IMF}$.  

Perhaps an effect such as strangulation \citep{Peng2015} would be more likely to produce the observed lower $T_{IMF}$ in the last stages of star formation.  In such a scenario, increasingly cooler gas would form stars even more efficiently, eventually leading to the depletion of the remaining gas and quiescence.  

\subsection{Synchronized Turnoff and Implications for Uniform Models}
\label{subsec:turnoff}

Other possible mechanisms are difficult to reconcile with the apparent universality of the drop in $T_{IMF}$.  It appears that nearly every quiescent galaxy exhibits a lower $T_{IMF}$, meaning that they were all in a similar state while quenching.  Thus, it is likely that most galaxies quenched via a common mechanism.  

A stronger form of this universality of quenching comes from looking at the quiescent fraction of galaxies as a function of mass and redshift.  Although nearly every stellar mass in this work is lower than the one inferred for the same galaxy using a Galactic IMF, the masses of star-forming galaxies decrease more than quiescent galaxies, since $T_{IMF}$ is generally higher.  As a result, previous estimates of quiescent fractions are replaced by a far simple picture: at every redshift, nearly all of the higher-mass galaxies are quiescent and nearly all of the lower-mass galaxies are star-forming.  There is a narrow mass range in which galaxies are transitioning from star-forming to quiescent, and that mass range monotonically decreases towards lower redshift. 

A key implication is that not only do all galaxies eventually become quiescent, but galaxies at the same stellar mass quench nearly all at the same time.  This would appear to be inconsistent with proposed quenching mechanisms such as galaxy harrassment \citep{Cortese2006,Wetzel2013,Woo2015,Bluck2019} and galaxy mergers or resulting AGN heating \citep{Mihos1996,DiMatteo2005}, which would be driven by local environmental differences.  

Instead, it points at a story much more similar to that of the star-forming main sequence, which is similarly synchronous \citep{Kelson2014,Steinhardt2016}.  Indeed, in Paper II, it is shown that $T_{IMF}$ also decreases moving upward in mass along the star-forming main sequence.  Thus, perhaps quenching is simply the natural endpoint of the same astrophysics and feedback mechanisms that drive the star-forming main sequence.  This, again, would appear to be more consistent with strangulation \citep{Peng2015,Henriques2019} or some other gradual depletion mechanism.  At a minimum, it is another strong indication that quenching, like many other things in galaxy evolution, is driven by processes which apply to nearly all galaxies in similar ways, rather than by a stochastic, environmentally-driven quenching mechanism.  

\section{Acknowledgements}

The authors would like to thank Vasily Kokorev and Darach Watson for useful discussions.  CLS is supported by ERC grant 648179 "ConTExt".  BM is supported by the Tombrello Fellowship. AL is supported by the Selove Prize.  The Cosmic Dawn Center (DAWN) is funded by the Danish National Research Foundation under grant No. 140.  The Flatiron Institute is supported by the Simons Foundation.

\bibliographystyle{mnras}
\bibliography{refs.bib} 

\label{lastpage}
\end{document}